\documentclass[%
 aip,
 amsmath,amssymb,
 reprint,%
]{revtex4-1}

\usepackage{graphicx}
\usepackage{subfig}
\usepackage{dcolumn}
\usepackage{bm}

\usepackage[utf8]{inputenc}
\usepackage[T1]{fontenc}
\usepackage{mathptmx}
\usepackage{etoolbox}

\usepackage{enumitem}

\makeatletter
\def\@email#1#2{%
 \endgroup
 \patchcmd{\titleblock@produce}
  {\frontmatter@RRAPformat}
  {\frontmatter@RRAPformat{\produce@RRAP{*#1\href{mailto:#2}{#2}}}\frontmatter@RRAPformat}
  {}{}
}%
\makeatother
\begin{document}

\preprint{AIP/123-QED}

\title[4-momentum conservation as the principal framework for mesonic decay]{4-momentum conservation as the principal framework for mesonic decay: The case of helium-5-lambda}
\author{Emile Meoto}
 \email{meoto.emile@ubuea.cm}
 \affiliation{Department of Physics, University of Buea, P O Box 63 Buea, South West Region, Cameroon.}
\author{Mantile L. Lekala}%
 \email{lekalml@unisa.ac.za}
\affiliation{Department of Physics, University of South Africa, Christiaan de Wet Road \& Pioneer Avenue Johannesburg, 1709 Gauteng, South Africa}%

\date{\today}

\begin{abstract}
Negative-pion and neutral-pion mesonic decays of the hypernucleus $ _\Lambda ^5$He are investigated within two-body and three-body relativistic kinematics in the rest-frame of the hypernucleus. 4-momentum conservation fully constrains the two-body decay, giving rise to a monochromatic pion momentum that is determined through the Newton-Raphson root-finding algorithm. In the case of three-body decay, where 4-momentum conservation is insufficient, the final-state momenta are parametrised by variables whose variations account for all kinematically allowed momenta. A Monte Carlo sampling method is then employed to generate 50 000 decay events per channel that simultaneously satisfy both energy and momentum conservation. The resulting pion momentum distributions exhibit clear peaks at 103.0 MeV/c for neutral pion decay and 97.3 MeV/c for negative pion decay. In addition, only 0.27\% of neutral pion decay events and 0.03\% of negative pion decay events produce nucleons with momenta exceeding the Fermi momentum of $  ^4$He, and are therefore allowed by the Pauli exclusion principle.
\end{abstract}

\maketitle

\section{Introduction}

The in-medium mesonic decay of a lambda hyperon provides a unique view on the role that hyperons play inside hypernuclei. In particular, it provides insights into the weak interaction in the baryonic sector, especially on strangeness-changing processes. In free space, the lambda hyperon undergoes weak decay into a pion and a nucleon. In a nuclear medium, this decay is modified due to the surrounding nuclear environment, which shifts the Q-values, and hence the phase space and kinematic variables of the final-state particles. The daughter nucleus may be stable or unstable. If it is stable, then the process is a two-body decay. If it is unstable (a resonance), then it further breaks up, in which case the process becomes a three-body decay. The neutron or proton produced in the decay must enter available nuclear states, where low-momentum states are already occupied up to the Fermi momentum. If this decay neutron or proton emerges with a momentum less than this Fermi momentum, it is Pauli blocked. Therefore, even if the decay is energetically favourable, it may still be Pauli blocked.  Two-body decay pion spectroscopy may be used in identifying new hypernuclei with very high accuracy \cite{nagao2023, maj2006}. For example, the observation of $_\Lambda ^4\text{H}$ (Ref. \cite{ess2015}) and recently $_\Lambda ^3\text{H}$ (Ref. \cite{ach2022}).

The distribution of momenta of final-state particles in mesonic decay is predominantly determined by four key factors: relativistic kinematics, Pauli blocking, Fermi motion of lambda hyperon and final-state interactions. The goal of this paper is to show that relativistic kinematics serves as the principal framework in the determination of these final-state momenta. This is illustrated through pion mesonic decay of the hypernucleus $_\Lambda ^5\text{He}$. Given that weak decays in emulsion experiments were the most widely used observational methods for hypernuclei, the mesonic weak decay of this hypernucleus received significant attention in those early years \cite{hil1958, tan1958, szy1958, dal1959, bye1959, amm1959, har1968, gaj1969}.

\section{Free lambda mesonic decay}

In free space, the dominant mesonic decay modes of the \(\Lambda\) hyperon are negative-pion and neutral-pion decays.

\begin{align}
\Lambda &\to p + \pi^- \\
\Lambda &\to n + \pi^0
\end{align}

These decays have Q-values given by

\begin{align}
Q_{\pi^-} = [m_\Lambda - m_p - m_{\pi^-}]c^2 \\
Q_{\pi^0} = [m_\Lambda - m_n - m_{\pi^0} ]c^2
\end{align}

where $m_pc^2$, $m_{\pi^-}c^2$, $m_nc^2$, and $m_{\pi^0}c^2 $ are tabulated masses (in units of MeV) of the proton, negative pion, neutron and neutral pion, respectively. The Q-values for these free decays are therefore given by

\begin{align}
Q_{\pi^-} &= [m_\Lambda - m_p - m_{\pi^-}]c^2 \notag  \\
      &= 1115.683 - 938.272 - 139.570 = 37.841 MeV \\
Q_{\pi^0} &= [m_\Lambda - m_n - m_{\pi^0} ]c^2 \notag \\
      &= 1115.683 - 939.565 - 134.977 = 41.141 MeV 
\end{align}

\section{In-medium mesonic decay}

Inside a hypernucleus $_{\Lambda}^{A} Z$, mesonic decays take place as follows:

\begin{align}\label{eq:decay_general_form}
_{\Lambda}^{A} Z &\to \ ^{A}(Z+1) + \pi^- \\
_{\Lambda}^{A} Z &\to \ ^{A}Z + \pi^0
\end{align}

The effective Q-values for these mesonic decays are modified by the nuclear environment. From standard definitions, these effective values are given by the following relations:

\begin{subequations}
\begin{align}\label{eq:proton_eff} 
Q_{\pi^-}^{eff} &= [ M(A,Z)_\Lambda -  M(A,Z+1) - m_{\pi^-}]c^2 \\
\label{eq:neutron_eff}
Q_{\pi^0}^{eff} &= [ M(A,Z)_\Lambda -  M(A,Z) - m_{\pi^0} ]c^2
\end{align} 
\end{subequations}

In order to illustrate how the nuclear medium modifies the free lambda Q values, these effective Q-values are rewritten in a different approach through the lambda separation energy. This is done first in the case of negative-pion decay. The lambda separation energy of the hypernucleus and the proton separation energy of the nucleus after decay are given by

\begin{subequations}
 \begin{align}
\label{eq:lambda}
B_\Lambda &= B(A, Z)_{\Lambda} - B(A-1, Z) \\
\label{eq:proton}
S_p &=  B(A, Z+1) - B(A-1,Z)
\end{align}   
\end{subequations}

Using the mass defect, these may be written as 

\begin{align}
B(A, Z)_\Lambda &= \left[ Z m_p + (A - Z - 1) m_n + m_\Lambda - M(A, Z)_\Lambda \right] c^2 \\
B(A, Z) &= \left[ Z m_p + (A - Z) m_n - M(A, Z) \right] c^2 \\
B(A-1, Z) &= \left[ Z m_p + (A-1 - Z) m_n - M(A-1, Z) \right] c^2 \\
B(A, Z+1) &= \left[ (Z+1)m_p + (A - (Z+1))m_n - M(A, Z+1) \right] c^2 \\
B(A-1,Z) &= \left[ Zm_p + (A-1 - Z)m_n - M(A-1,Z) \right] c^2
\end{align}

Substituting these binding energy expressions into Eq. \ref{eq:lambda} to find $B_\Lambda $ and simplifying \cite{bur1965}

\begin{align}
B_\Lambda &= \left( \left[ Zm_p + (A-Z-1)m_n + m_\Lambda - M(A,Z)_\Lambda \right] c^2 \right) \notag \\
&\quad - \left( \left[ Zm_p + (A-Z-1)m_n - M(A-1,Z) \right] c^2 \right) \notag \\
&= \left[ Zm_p + (A-Z-1)m_n + m_\Lambda - M(A,Z)_\Lambda \right. \notag \\
&\quad \left. - Zm_p - (A-Z-1)m_n + M(A-1,Z) \right] c^2 \\
\rightarrow B_\Lambda &= \left[ M(A-1,Z) + m_\Lambda - M(A,Z)_\Lambda \right] c^2
\end{align}

Similarly, binding energies are substituted in Eq. \ref{eq:proton} to find $S_p$ in terms of masses.

\begin{align}
S_p &= \Big[ (Z+1)m_p + (A-Z-1)m_n - M(A, Z+1) \Big] c^2 \notag \\
&\quad - \Big[ Zm_p + (A-1-Z)m_n - M(A-1, Z) \Big] c^2 \notag \\
&= \Big[ Zm_p + m_p + (A-Z-1)m_n - M(A, Z+1) \notag \\
&\quad - Zm_p - (A-Z-1)m_n + m_n + M(A-1, Z) \Big] c^2 \notag \\
&= \left[ m_p + M(A-1, Z) - M(A, Z+1) \right] c^2
\end{align}

Therefore, the lambda binding energy and proton separation energy are given by
\begin{align}
B_\Lambda &= \left[  M(A-1,Z) + m_\Lambda - M(A,Z)_\Lambda \right] c^2 \\
S_p &= \left[ m_p + M(A-1, Z) - M(A, Z+1) \right] c^2 
\end{align}

From these equations,
\begin{align}
M(A,Z)_\Lambda&= M(A-1,Z)  + m_\Lambda - \frac{B_\Lambda}{c^2} \\
M(A, Z+1) &= M(A-1, Z) + m_p - \frac{S_p}{c^2}
\end{align}

Substituting these masses into the expression for \( Q_{\pi^-}^{\text{eff}} \) (Eq.\ref{eq:proton_eff}):

\begin{eqnarray}
Q_{\pi^-}^{\text{eff}} &&= \left[ \left( M(A-1,Z)  + m_\Lambda - \frac{B_\Lambda}{c^2} \right) \right. \notag \\
&& \left.- \left( M(A-1, Z) + m_p - \frac{S_p}{c^2} \right) - m_{\pi^-} \right] c^2 \notag \\
&&= \left[ m_\Lambda - m_p - m_{\pi^-} - \frac{B_\Lambda}{c^2} + \frac{S_p}{c^2} \right] c^2 \notag \\
&&= \left( m_\Lambda - m_p - m_{\pi^-} \right) c^2 - B_\Lambda + S_p \notag \\
&&= Q^p - B_\Lambda + S_p
\end{eqnarray}

Therefore, the in-medium Q-value expressed as a correction to the free Q-value is \cite{bur1965, alb2002}

\begin{align}\label{eq:qeff_charged}
Q_{\pi^-}^{\text{eff}} = Q_{\pi^-} - B_\Lambda + S_p
\end{align}
Following a similar procedure and using the neutron separation energy for the daughter nucleus ($S_n=[m_n + M(A-1, Z) - M(A, Z)]c^2$), it may be shown that 

\begin{align}\label{eq:qeff_neutral}
Q_{\pi^0}^{\text{eff}} = Q_{\pi^0} - B_\Lambda + S_n
\end{align}

As it has been shown, the Q-value of these competing processes are determined solely by $B_\Lambda$, $S_p$ and $S_n$. The expressions in Eqs. \ref{eq:qeff_charged} and \ref{eq:qeff_neutral} are very useful because they involve the quantities $B_\Lambda$, $S_p$ and $S_n$ that can be measured in an experiment.

\section{Relativistic kinematics}

Consider the two-body mesonic decay $X \to Y + \pi$, where $\pi$ represents the negative pion or the neutral pion. In the rest-frame of X, let the 4-momenta of all particles be as follows:

\begin{align}
X &:P_{X}^\mu= \left(\frac{E_{X}}{c}, \vec{p}_{X} \right)= \left(M(X)c, \vec{0} \right) \\
Y &: P_{Y}^\mu = \left(\frac{E_{Y} }{c}, \vec{p}_{Y} \right) \\
\pi &: P_{\pi}^\mu = \left(\frac{E_{\pi} }{c}, \vec{p}_{\pi} \right) 
\end{align}

Assuming the system is isolated, conservation of 4-momentum is given by

\begin{align}
P_{X}^\mu = P_{Y}^\mu + P_{\pi}^\mu 
\end{align}

This equation is equivalent to energy conservation and momentum conservation:
\begin{align}
M(X)c^2 = E_{Y} + E_{\pi} \\
\vec{0}= \vec{p}_{Y} + \vec{p}_{\pi} 
\end{align}

where $E_i=\gamma m_i c^2$ is the relativistic energy, $\vec{p}_{i}=\gamma m_i \vec{v}_i$ is the relativistic 3-momentum of particle $i$ and $\vec{v}_i$ is its velocity. $\gamma$ is the Lorentz factor. For each body, the Minkowski norm of the 4-momentum (a Lorentz invariant) is given by $P_{i}^\mu (P_{i})_\mu = \eta_{\mu \nu} P_{i}^\mu P_{i}^\nu = -m^2_ic^2 = -(E/c)^2 + |\vec{p}_i|^2 $, where $\eta_{\mu \nu} = \text{diag}(-1,+1,+1,+1)$ is the Minkowski metric tensor and $m_i$ is the invariant mass of particle $i$. Energy conservation may therefore be written as follows:

\begin{align}
M(X)c^2 = \sqrt{(m_{Y}c^2)^2 + |\vec{p}_{Y}|^2c^2} + \sqrt{(m_{\pi}c^2)^2 + |\vec{p}_{\pi}|^2c^2}
\end{align}

Energy conservation is rewritten in terms of the Q-value, as follows:

\begin{align}
& M(X)c^2 - m_{Y}c^2 - m_{\pi}c^2 = \\
& \sqrt{(m_{Y}c^2)^2 + |\vec{p}_{Y}|^2c^2} - m_{Y}c^2 + \sqrt{(m_{\pi}c^2)^2 + |\vec{p}_{\pi}|^2c^2} - m_{\pi}c^2
\end{align}

The left-hand side of this equation may be recognised as the effective Q-value of the mesonic decay while the right-hand side is the sum of the kinetic energies of the products of the decay. Therefore, the equations for energy conservation and momentum conservation may be written as 

\begin{align}
\label{eq:general_two_energy}
Q_{\pi}^{\text{eff}} = T_Y + T_{\pi} \\
\label{eq:general_two_momentum}
\vec{0}= \vec{p}_{Y} + \vec{p}_{\pi} 
\end{align}

This form of the equation clearly illustrates that the Q-value ($Q_{\pi}^{\text{eff}}$) provides the energy that is partitioned into the kinetic energies of X and the pion. In the case of the three-body decay $X \to Y + Z + \pi$, conservation of 4-momentum leads to the following equations for energy conservation and momentum conservation 

\begin{align}
\label{eq:general_three}
Q_{\pi}^{\text{eff}} = T_Y + T_Z + T_{\pi} \\
\vec{0}= \vec{p}_{Y} + \vec{p}_{Z} + \vec{p}_{\pi} 
\end{align}

Unlike the case of two-body decay, where momentum conservation requires the two final-state particles to be emitted back-to-back (collinear and oppositely directed), the three-body decay imposes no such collinearity constraint. The three momenta can assume any relative directions, only subject to conservation. This leaves an infinite number of kinematically allowed angular configurations.
\section{Results and Discussion}

In $_\Lambda ^5\text{He}$, negative pion and neutral pion mesonic decays are represented by the nuclear equations

\begin{align}
_{\Lambda}^{5}\text{He} &\to \ ^{5}\text{Li} + \pi^- \\
_{\Lambda}^{5} \text{He} &\to \ ^{5}\text{He} + \pi^0
\end{align}

Equations \ref{eq:qeff_charged} and \ref{eq:qeff_neutral} are used to compute the in-medium Q-values. The experimental lambda binding energy used is $B_\Lambda=3.12$ MeV, while the separation energies $S_p(^{5}\text{Li})=-1.960$ MeV and $S_n(^{5}\text{He})=-0.735$ MeV are recorded values from Ref. \cite{wan2021}. These effective Q-values are

\begin{align}
Q_{\pi^-}^{\text{eff}} = 37.841 - 3.12 -1.960 = 32.76 \,\, \text{MeV} \\
Q_{\pi^0}^{\text{eff}} = 41.141 - 3.12 -0.735 = 37.29 \,\, \text{MeV}
\end{align}

Inside the hypernucleus, negative-pion mesonic decay competes with neutral-pion mesonic decay. From these values, two conclusions are evident: (i) Since $Q_{\pi^-}^{\text{eff}} >0$ and $Q_{\pi^0}^{\text{eff}} >0$, both negative pion and neutral pion mesonic decays are energetically allowed in $_{\Lambda}^{5} \text{He}$. (ii) Neutral pion mesonic decay is more energetically favourable than negative pion mesonic decay since $Q_{\pi^0}^{\text{eff}} > Q_{\pi^-}^{\text{eff}}$. In the following section, the phase space of each decay is determined through two-body and three-body relativistic kinematics in the rest frame of $_{\Lambda}^{5}\text{He}$.
\subsection{Two-body relativistic kinematics of mesonic decay}

\subsubsection{Negative pion decay}

$_{\Lambda}^{5}\text{He}$ undergoes negative-pion mesonic decay into a $^{5}\text{Li}$ daughter nucleus and a $\pi^-$ (Eq. \ref{eq:decay_general_form}).
\begin{align}
_{\Lambda}^{5}\text{He} &\to \ ^{5}\text{Li} + \pi^- \\
\end{align}

From Eqs. \ref{eq:general_two_energy} and \ref{eq:general_two_momentum}, 
\begin{align}
Q_{\pi^-}^{\text{eff}} = T_{5Li} + T_{\pi^-} \\
\vec{0}= \vec{p}_{5Li} + \vec{p}_{\pi^-} 
\end{align}
where 
\begin{align}
T_{5 Li} &= \sqrt{(m_{5 Li}c^2)^2 + (p_{\pi^-}c)^2} - m_{5 Li}c^2\\
T_{\pi^-}&= \sqrt{(m_{\pi^-}c^2)^2 + (p_{\pi^-}c)^2} - m_{\pi^-}c^2
\end{align}

Using momentum conservation, $\vec{p}_{5Li}$ may be written in terms of $\vec{p}_{\pi^-}$ ($\vec{p}_{5Li} =- \vec{p}_{\pi^-} $), and the energy conservation equation is now expressed in terms of $\vec{p}_{\pi^-}$. Substituting $ Q_{\pi^-}^{\text{eff}}=32.76$ MeV, $m_{5 Li}c^2=4669.071$ MeV, and $m_{\pi^-}c^2=139.57039$ MeV, this equation for $\vec{p}_{\pi^-}$ is given by

\begin{align}
    32.76 = \sqrt{4669.071^2 + p_{\pi^-}^2c^2} - 4669.071 \notag \\
    + \sqrt{139.57039^2 +p_{\pi^-}^2c^2} - 139.57039
\end{align}

where. $|\vec{p}_{\pi^-}|=p_{\pi^-}$. Let the function $f(p_{\pi^-})$ be defined as

\begin{align}
f(p_{\pi^-}) &= \sqrt{4669.071^2 + p_{\pi^-}^2c^2} \\
&+ \sqrt{139.57039^2 + p_{\pi^-}^2c^2} -4841.40139 
\end{align}

The equation $f(p_{\pi^-})=0$ is then solved for the negative-pion momentum. This solution may be found through algebraic manipulations i.e. squaring repeatedly to remove the square root. However, this method would be ill-advised because it leads to high-degree polynomials that are difficult to solve. A better approach would be to use a numerical root-finding algorithm. In this paper, the Newton-Raphson algorithm is used. The first derivative used in this algorithm is 

\begin{align}
f'(p_{\pi^-}) = \frac{p_{\pi^-}c^2}{\sqrt{4669.071^2 + p_{\pi^-}^2c^2}} + \frac{p_{\pi^-}c^2}{\sqrt{139.57039^2 + p_{\pi^-}^2c^2}}    
\end{align}

Starting with an initial guess of 50 MeV/c, the root was found at $p_{\pi^-}=99.27$ MeV/c after 5 iterations. The collinear momentum of $^5$Li is $p_{5Li}=99.27$ MeV/c, in a direction opposite to that of the negative-pion. The kinetic energy of each body is given by

\begin{align}
T_{5 Li} &= \sqrt{(m_{5 Li}c^2)^2 + (p_{\pi^-}c)^2} - m_{5 Li}c^2=1.06 MeV\\
T_{\pi^-}&= \sqrt{(m_{\pi^-}c^2)^2 + (p_{\pi^-}c)^2} - m_{\pi^-}c^2 = 31.70 MeV
\end{align}
This shows that from the total available energy ($ Q_{\pi^-}^{\text{eff}}=32.76$ MeV), the kinetic energy given to the negative pion is 31.70 Mev (96.76 \% of $ Q_{\pi^-}^{\text{eff}}$),  while a very small fraction of 1.06 MeV (3.24 \% of $ Q_{\pi^-}^{\text{eff}}$) is the recoil kinetic energy of the $^5$Li. 
\subsubsection{Neutral pion decay}

Neutral pion decay in $_{\Lambda}^{5} \text{He}$ results in a $^{5}\text{He}$ daughter nucleus and a $\pi^0$.

\begin{align}
_{\Lambda}^{5} \text{He} &\to \ ^{5}\text{He} + \pi^0
\end{align}
Following a similar procedure, and using the constraint from momentum conservation $\vec{p}_{5He} =- \vec{p}_{\pi^0}$, the equation for $p_{\pi^0}$ is given by

\begin{align}
Q_{\pi^0}^{\text{eff}} &= \sqrt{(m_{5 He}c^2)^2 + (-p_{\pi^0}c)^2} - m_{5 He}c^2  \notag \\
 &+ \sqrt{(m_{\pi^0}c^2)^2 + (p_{\pi^0}c)^2} - m_{\pi^0}c^2
\end{align}

Using $Q_{\pi^0}^{\text{eff}}=37.29$ MeV, together with the masses $m_{5 He}=4668.174$ MeV/$c^2$ and $m_{\pi^0}=134.9768$ MeV/$c^2$, this equation becomes 

\begin{align}
37.29 &= \sqrt{4668.174^2 + p_{\pi^0}^2c^2} - 4668.174 \notag \\
&+ \sqrt{134.9768^2 + p_{\pi^0}^2c^2} - 134.9768 \\
&\sqrt{4668.174^2 + p_{\pi^0}^2c^2}  \notag \\
&+ \sqrt{134.9768^2 + p_{\pi^0}^2c^2} - 4840.4408=0
\end{align}

Through the Newton-Raphson numerical scheme, the root $p_{\pi^0}=105.12$ MeV/c was found. From momentum conservation, the recoil momentum of 5He is $p_{5He} =105.12$ MeV. The kinetic energy of each particle is given by

\begin{align}
T_{5 He} &= \sqrt{(m_{5 He}c^2)^2 + (p_{\pi^0}c)^2} - m_{5 He}c^2 = 1.19 MeV\\
T_{\pi^0}&= \sqrt{(m_{\pi^0}c^2)^2 + (p_{\pi^0}c)^2} - m_{\pi^0}c^2 = 36.11 MeV
\end{align}

It may be observed that from the energy released, $Q_{\pi^0}^{\text{eff}}=37.29$ MeV, the kinetic energy given to the neutral pion is 36.11 MeV (96.84 \% of $Q_{\pi^0}^{\text{eff}}$) while the kinetic energy given to $^5$He is 1.19 MeV (3.19 \%  of $Q_{\pi^0}^{\text{eff}}$). This is the recoil energy of $^5$He.

\subsection{Three-body relativistic kinematics of mesonic decay}

In the foregoing discussion, charged and neutral-pion mesonic decay in $_{\Lambda}^{5}\text{He}$ were treated as two-body decays. However, the daughter nuclei are not particle-bound. In fact, many pion mesonic decays result in unbound daughter nuclei, thereby leading to further fragmentation. From nuclear databases, one observes that $S_p(^{5}\text{Li})<0$ and $S_n(^{5}\text{He})<0$. This implies that in negative-pion mesonic decay of $_{\Lambda}^{5}\text{He}$, the $^5$Li nucleus further breaks up into $^4$He and a proton, while in neutral-pion decay $^5$He further breaks up into $^4$He and a neutron. In other words, $^5$He and $^5$Li are two-body resonances (they would show up as clusters in a Dalitz plot for the three-body decay). Using these final-state particles, the pion decay equations may be written as three-body decays.

\begin{subequations}
\begin{align}
\label{eq:charged_three}
_{\Lambda}^{5}\text{He} &\to \ ^{4}\text{He}+ p + \pi^- \\
\label{eq:neutral_three}
_{\Lambda}^{5} \text{He} &\to \ ^{4}\text{He}+ n + \pi^0
\end{align}  
\end{subequations}

Therefore, the total energy released, $Q_{\pi^-}^{\text{eff}}$, is partitioned into the kinetic energies of $^{4}\text{He}$, $p$ and $\pi^-$. Similarly, $Q_{\pi^0}^{\text{eff}}$ is partitioned into the kinetic energies of $^{4}\text{He}$, $n$ and $\pi^0$. These three-body decays are more complex than two-body decay, as will be illustrated. The case of neutral-pion decay is treated first. In the rest frame of $_{\Lambda}^{5}\text{He}$, let the 4-momentum of each particle be

\begin{align}
_{\Lambda}^{5}\text{He} &:P_{5 \Lambda He}^\mu= \left(\frac{E_{5 \Lambda He}}{c}, \vec{p}_{5 \Lambda He} \right)= \left(M(_{\Lambda}^{5}\text{He})c, \vec{0} \right) \\
^{4}\text{He} &: P_{4 He}^\mu = \left(\frac{E_{4 He} }{c}, \vec{p}_{4He} \right) \\
n &: P_{n}^\mu = \left(\frac{E_{n} }{c}, \vec{p}_{n} \right) \\
\pi^0 &: P_{\pi^0}^\mu = \left(\frac{E_{\pi^0} }{c}, \vec{p}_{\pi^0} \right) 
\end{align}

4-momentum conservation results in the following pair of equations

\begin{align}
\label{eq:energy_four}
Q_{\pi^0}^{\text{eff}} &= T_{4 He} + T_{n} + T_{\pi^0} \\
\label{eq:momentum_four}
\vec{0} &= \vec{p}_{4He} + \vec{p}_{n} + \vec{p}_{\pi^0} 
\end{align}

The energy conservation equation is given explicitly by

\begin{align}
Q_{\pi^0}^{\text{eff}} &= \sqrt{ |\vec{p}_{4 He}|^2c^2 + M(4 He)^2 c^4} - M(4 He) c^2 \\
&+ \sqrt{|\vec{p}_{n} |^2 c^2+ m_n^2 c^4} - m_n c^2 \notag \\
&+ \sqrt{|p_{\pi^0}|^2 c^2 + m_{\pi^0}^2 c^4} - m_{\pi^0} c^2
\end{align}

In the two-body decay problem, momentum conservation sufficiently constraints the kinematics. However, in the three-body case, momentum conservation allows for an infinite number of possible partitions. In this paper, an approach is proposed to find a \textit{sufficiently} large number of possible momentum partitions. Let the total momentum released by the decay be $p$. The momenta of $^4$He, the neutron and the neutral pion may be expressed as fractions of $p$.

\begin{align}
\vec{p}_{4He} &= q p \hat{n}_1 \\
\vec{p}_n &= r p \hat{n}_2 \\
\vec{p}_{\pi^0} &= s p \hat{n}_3,
\end{align}

where $\hat{n}_1$, $\hat{n}_2$ and $\hat{n}_3$ are unit vectors in random directions. In the strategy proposed here, two directions are chosen randomly, and the third is fixed by momentum conservation. In particular, the unit vectors $\hat{n}_1$ and $\hat{n}_2$ are chosen randomly, and the unit vector $\hat{n}_3$ is constrained by 

\begin{align}
\hat{n}_3 = -\frac{q \hat{n}_1 + r \hat{n}_2}{s}
\end{align}

The direction vectors $\hat{n}_1=(n_{1x}, n_{1y}, n_{1z})$ and $\hat{n}_2=(n_{2x}, n_{2y}, n_{2z})$ are parametrised as 3D vectors on the surface of a unit sphere. By randomly choosing directions for two of the three decay particles, the modelling of the decay is made as close to reality as possible. The azimuthal angles ($\phi_1$ and $\phi_2$) and the polar angles ($\theta_1$ and $\theta_2$) are randomly selected on the unit sphere, and components of the unit vectors (for $i=1,2$) are computed as

\begin{align}
n_{ix} &= \sin \theta_i \cos \phi_i \\
n_{iy} &= \sin \theta_i \sin \phi_i \\
n_{iz} &= \cos \theta_i
\end{align}

The energy conservation equation is given by

\begin{align}
Q_{\pi^0}^{\text{eff}} &= \sqrt{q^2p^2 c^2 + M(4 He)^2 c^4} \\
&- M(4 He) c^2 + \sqrt{r^2p^2 c^2 + m_n^2 c^4} - m_n c^2 \\
&+ \sqrt{s^2p^2 c^2 + m_{\pi^0}^2 c^4} - m_{\pi^0} c^2 
\end{align} 

Using $Q_{\pi^0}^{\text{eff}}=37.29$ MeV, $M(4 He)=3727.379$ MeV/$c^2$, $m_{\pi^0}=134.9768$ MeV/$c^2$ and $m_n=939.565$ MeV/$c^2$ one arrives at an equation for $p$.

\begin{align}
 37.29 &= \sqrt{(q^2p^2c^2  + (3727.379)^2} - 3727.379 \\
 &+ \sqrt{r^2p^2c^2 + (939.565)^2} - 939.565 \\
 &+ \sqrt{s^2p^2c^2 + (134.9768)^2} - 134.9768   
\end{align}

\begin{figure}[h]%
    \centering
    \subfloat[\centering {\small Neutral pion momentum distribution with a peak of 103.0 MeV/c}]{{\includegraphics[scale=0.2]{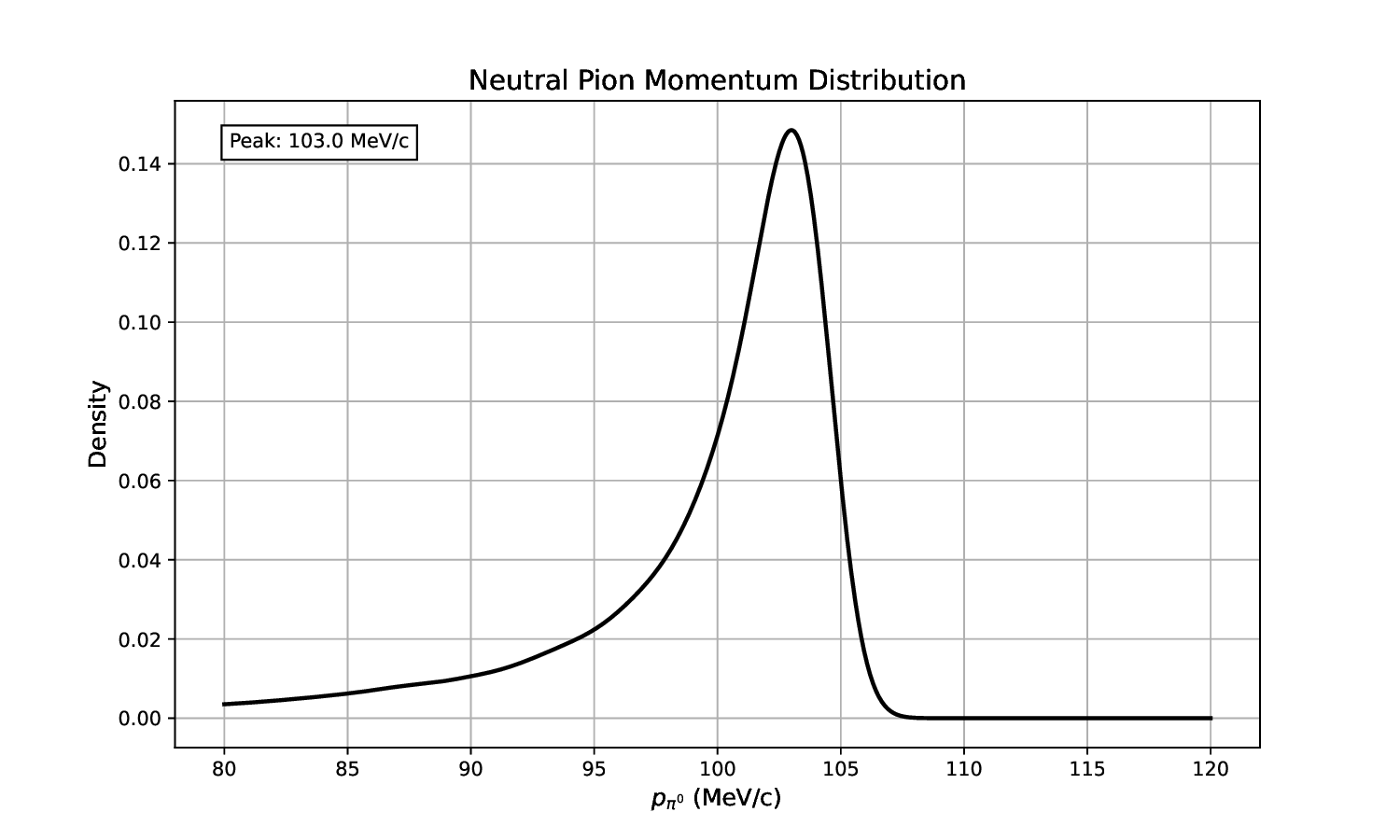} }}%
    \qquad
    \subfloat[\centering {\small Neutron and helium-4 momentum distributions. The peaks of these distributions are 61.9 MeV/c and 61.4 MeV/c, respectively.}]{{\includegraphics[scale=0.2]{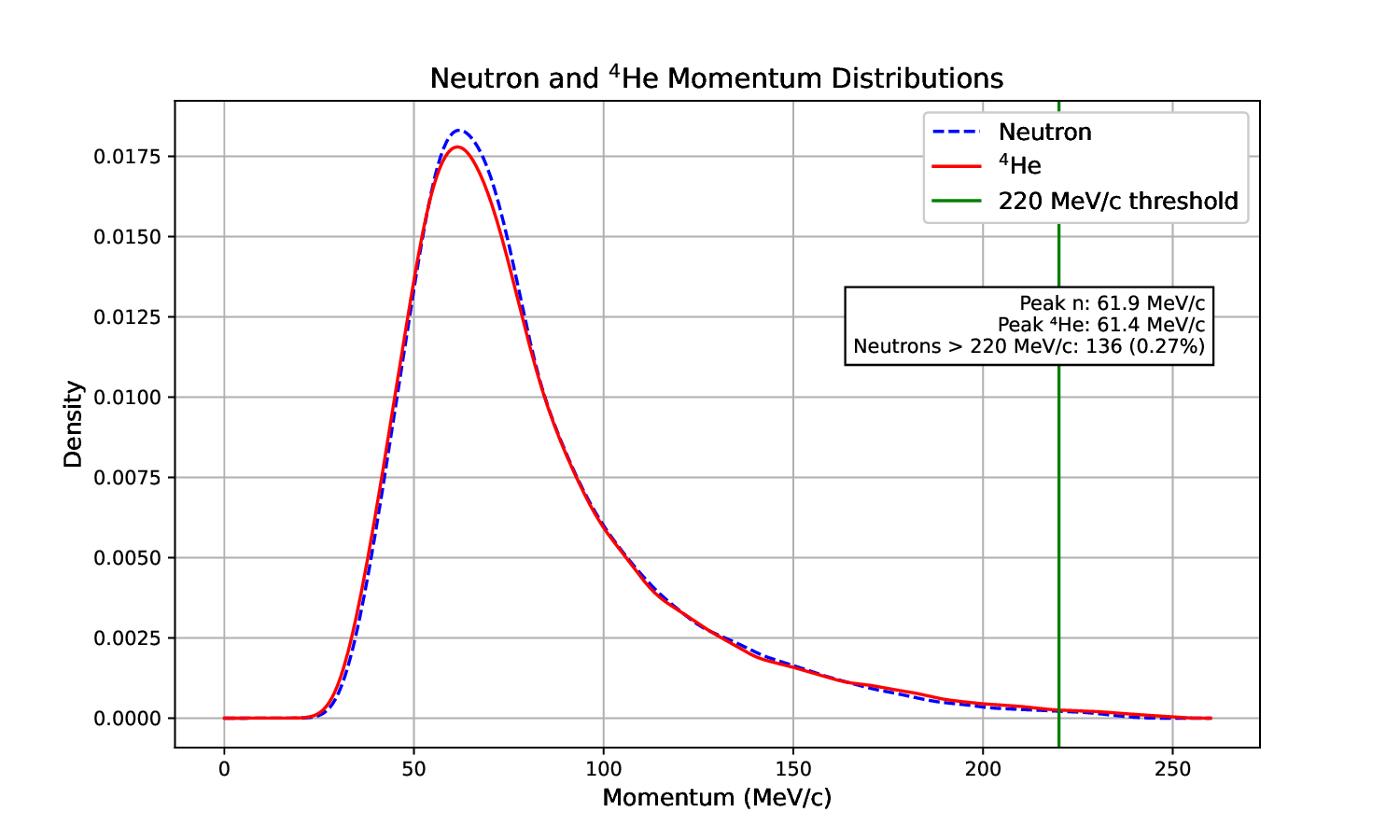} }}%

    \subfloat[\centering {\small Contour plot of neutron and helium-4 densities. This shows the kinematically allowed momentum region.}]{{\includegraphics[scale=0.25]{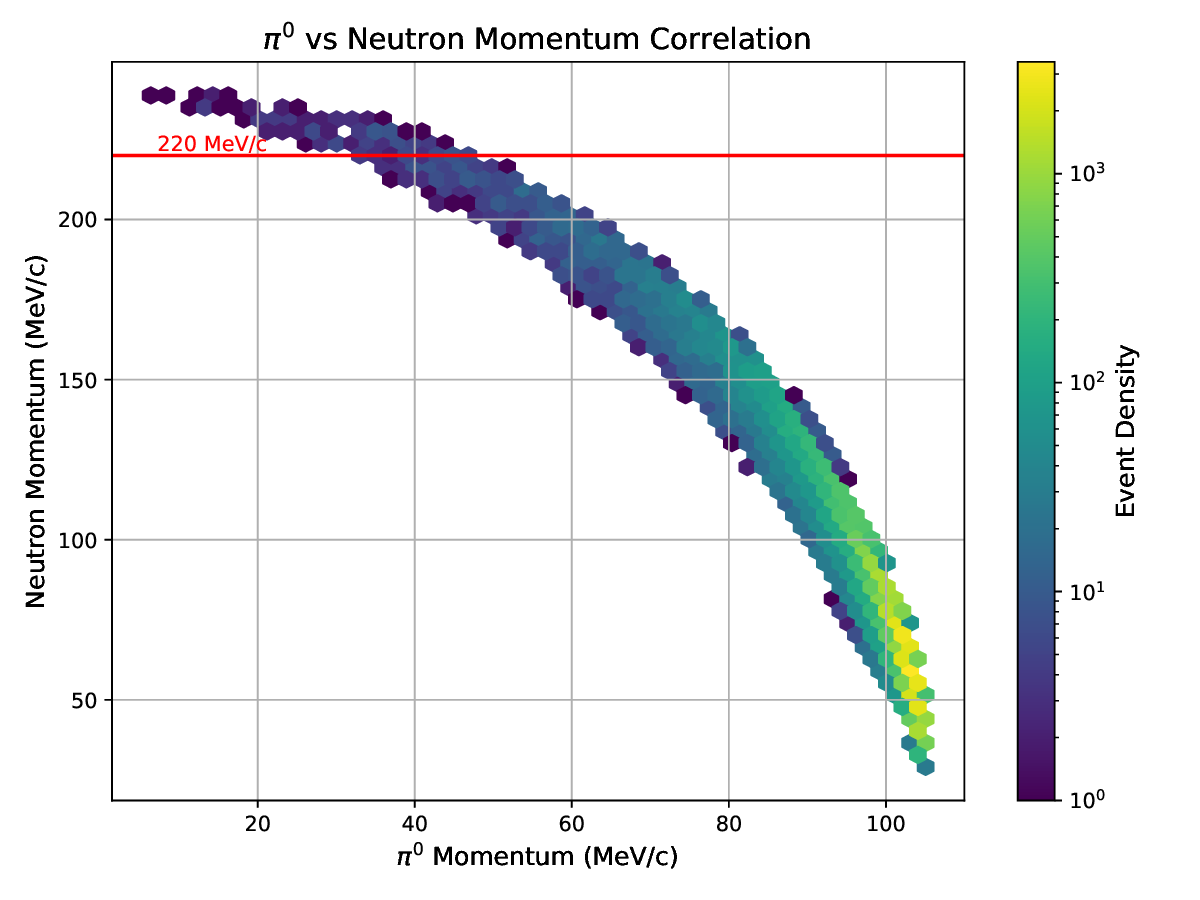} }}%
    \qquad
    \subfloat[\centering {\small Raw count (frequency) of the number of neutrons that fall within each momentum bin. The bin width is $(\text{max}(p_n) - \text{min}(p_n)/50$.}]{{\includegraphics[scale=0.25]{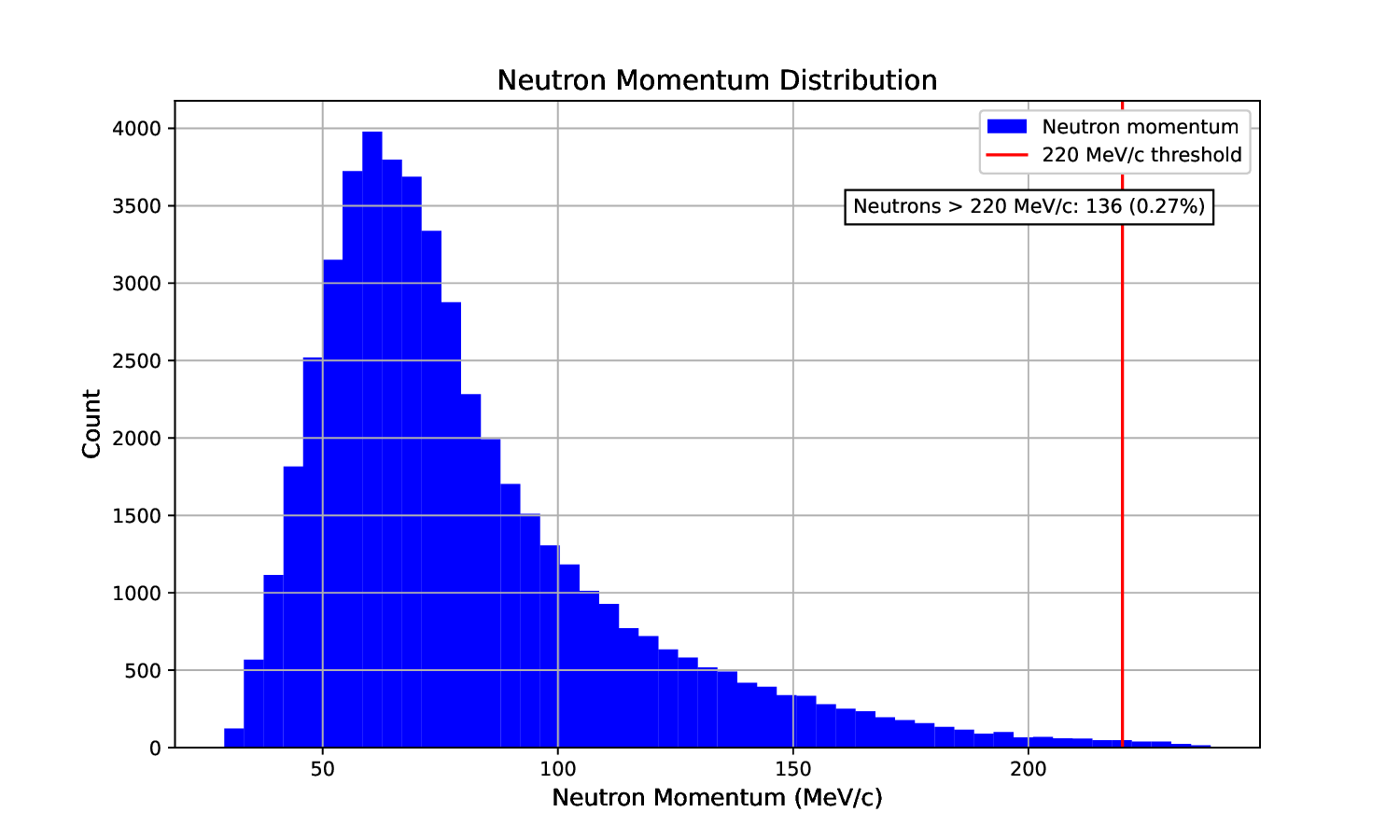} }}%
    \caption{Momentum distributions for neutral pion, neutron and helium-4 daughter nucleus in a three-body mesonic decays of $^5_{\Lambda} He$.}%
    \label{fig:neutral_plots}%
\end{figure}

There are many possible combinations of $q$, $r$, and $s$ that conform to 4-momentum conservation. In this paper, 50000 decay events are generated (for reproducibility, a seed of 42 is used) and a Monte Carlo approach is applied. For each decay event, $q$ and $r$ are randomly generated. Next, the directions $\hat{n}_1$ and $\hat{n}_2$ for $^4$He and neutron momenta are randomly generated. The momentum of the negative pion is calculated based on momentum conservation ($ \vec{p}_{\pi^0} = -(\vec{p}_{4He} + \vec{p}_{n})$). A root-finding algorithm is used to compute the momentum $p$ from the energy conservation equation. The events are checked to ensure that $p>0$ and $Q_{\pi^0}^{\text{eff}} = T_{4 He} + T_{n} + T_{\pi^0}$. The results from the application of this procedure are shown in Figure \ref{fig:neutral_plots}. This figure shows density of states for the neutral pion, density of states for $^4$He and the neutron, a contour plot of $^4$He and neutron momenta and a plot of the count of number of neutrons for each momentum bin.

For the negative pion mesonic decay, treated as a three-body decay, 4-momentum conservation results in 

\begin{align}
Q_{\pi^-}^{\text{eff}} &=  T_{4 He}  + T_p + T_{\pi^-} \\
\vec{0} &= \vec{p}_{4He} + \vec{p}_{p} + \vec{p}_{\pi^-} 
\end{align}

A similar momentum partitioning is used.

\begin{align}
\vec{p}_{4 He} &= q p \hat{n}_1 \\
\vec{p}_{p}    &= r p \hat{n}_2 \\
\vec{p}_{\pi^-} &=s p \hat{n}_3 
\end{align}

With $Q_{\pi^-}^{\text{eff}}=32.76$, and the masses $m_p=938.272$ MeV/$c^2$ and $m_{\pi^-}=139.5704 $ MeV/$c^2$, the energy conservation equation becomes

\begin{align}
Q_{\pi^-}^{\text{eff}} &=\sqrt{q^2p^2 c^2 + M(4 He)^2 c^4} - M(4 He) c^2 \notag \\
&+ \sqrt{r^2p^2 c^2 + m_p^2 c^4} - m_p c^2 \notag \\
&+ \sqrt{s^2p^2 c^2 + m_{\pi^-}^2 c^4} - m_{\pi^-} c^2  
\end{align} 

\begin{figure}[h!]%
    \centering
    \subfloat[\centering {\small Negative pion momentum distribution. The negative pion peak momentum is 97.3 MeV/c. This is the most probable momentum. }]{{\includegraphics[scale=0.2]{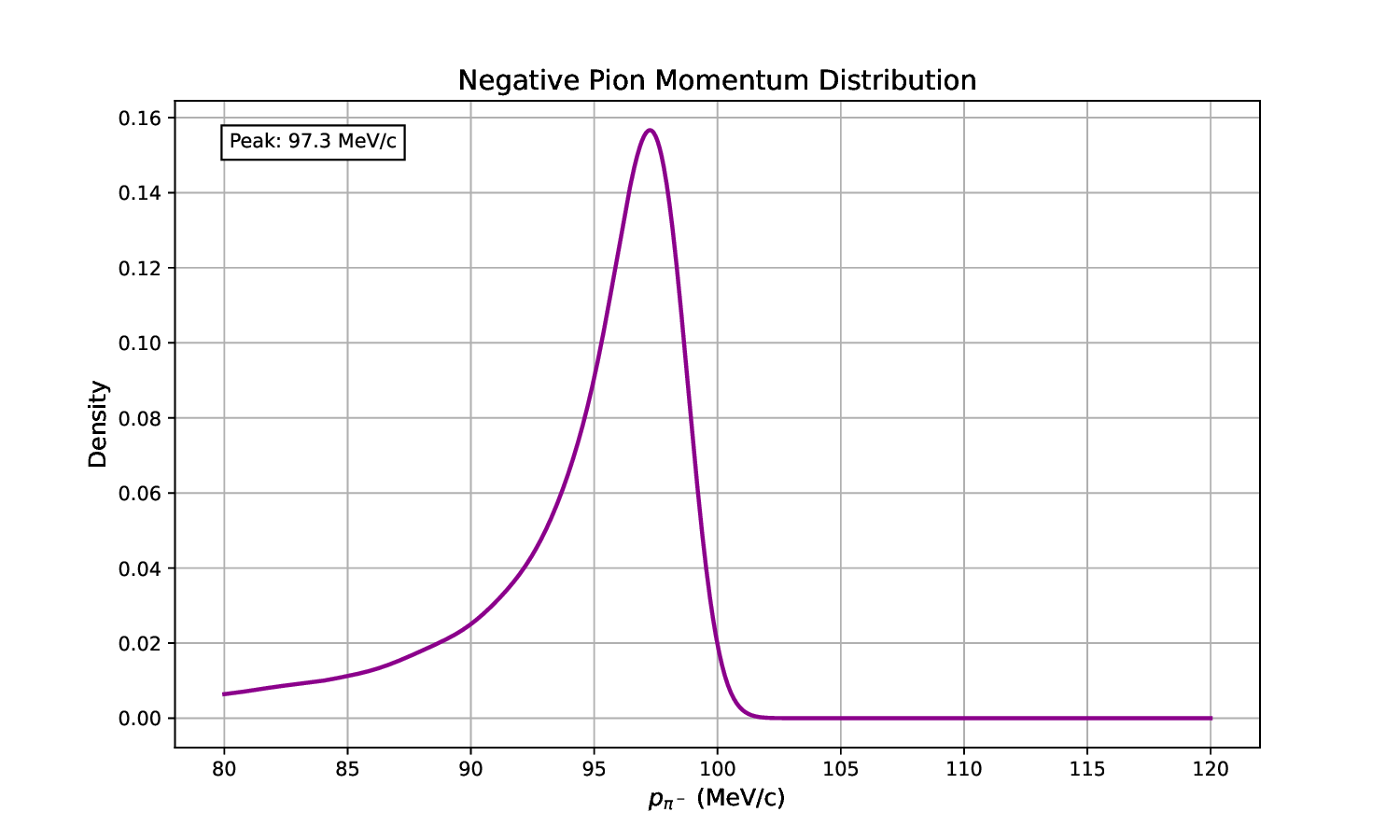} }}%
    \qquad
    \subfloat[\centering {\small Proton and helium-4 momentum distributions. The peaks for these distributions are 58.6 MeV/c and 58.0 MeV/c, respectively. }]{{\includegraphics[scale=0.2]{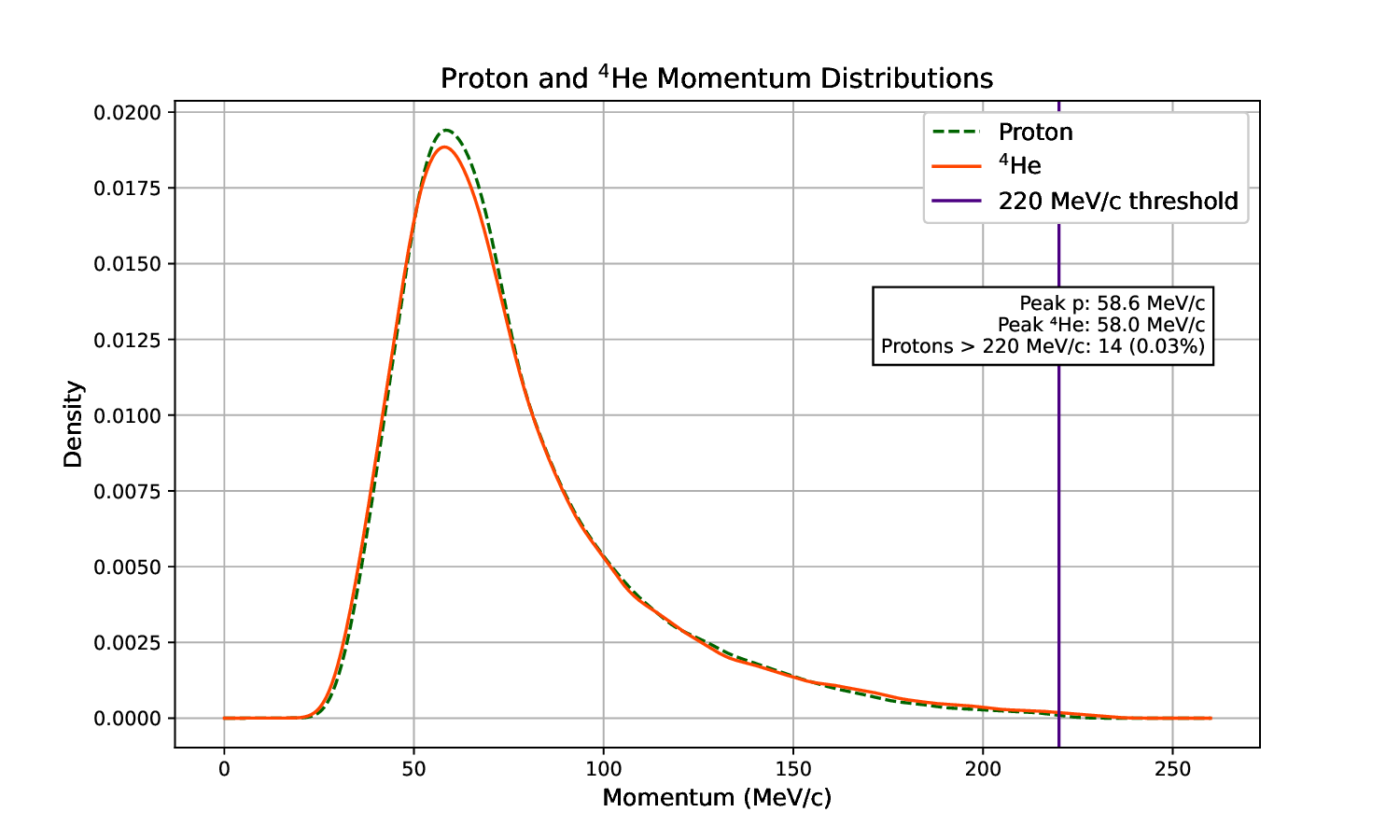} }}%

    \subfloat[\centering {\small Contour plot of proton and helium-4 densities. This shows the kinematically allowed momentum region. }]{{\includegraphics[scale=0.25]{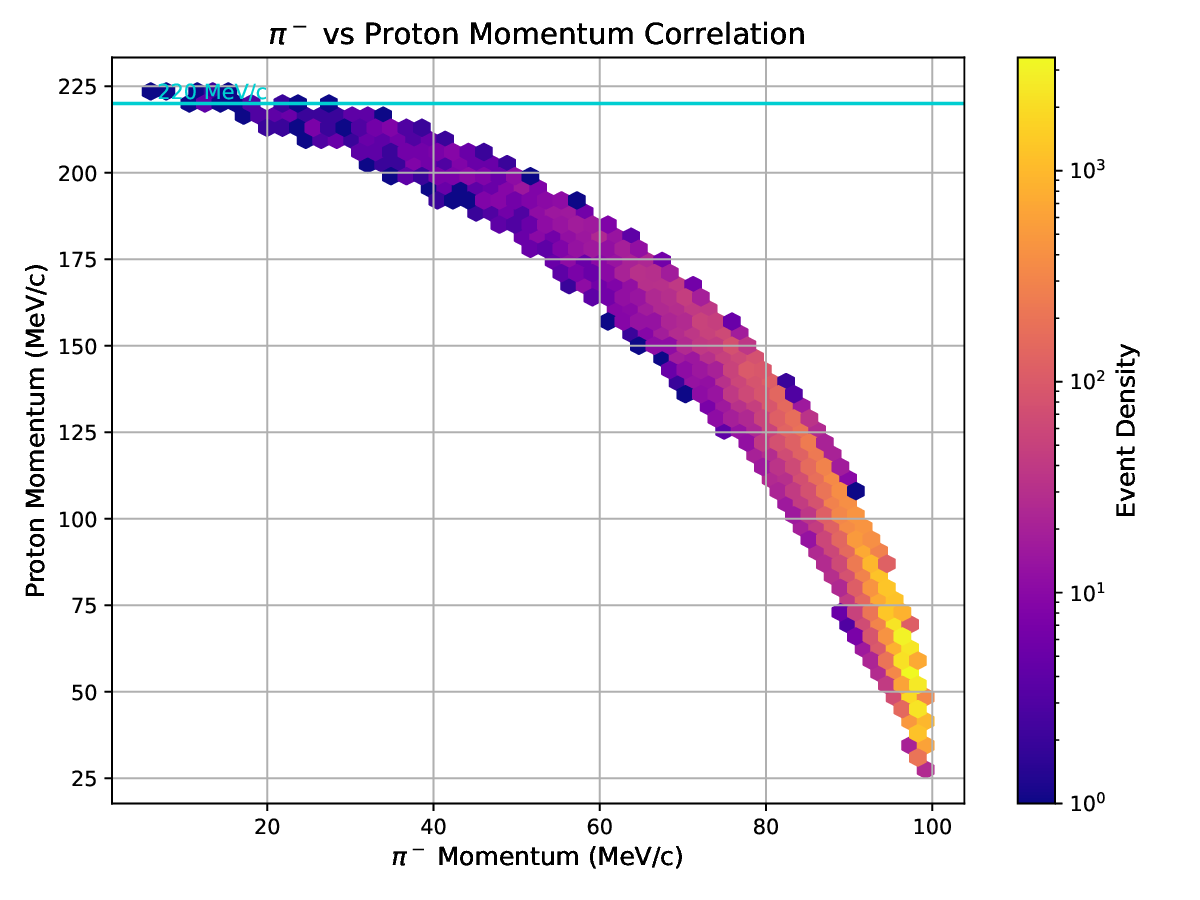} }}%
    \qquad
    \subfloat[\centering {\small Raw count (frequency) of the number of protons that fall within each momentum bin. }]{{\includegraphics[scale=0.25]{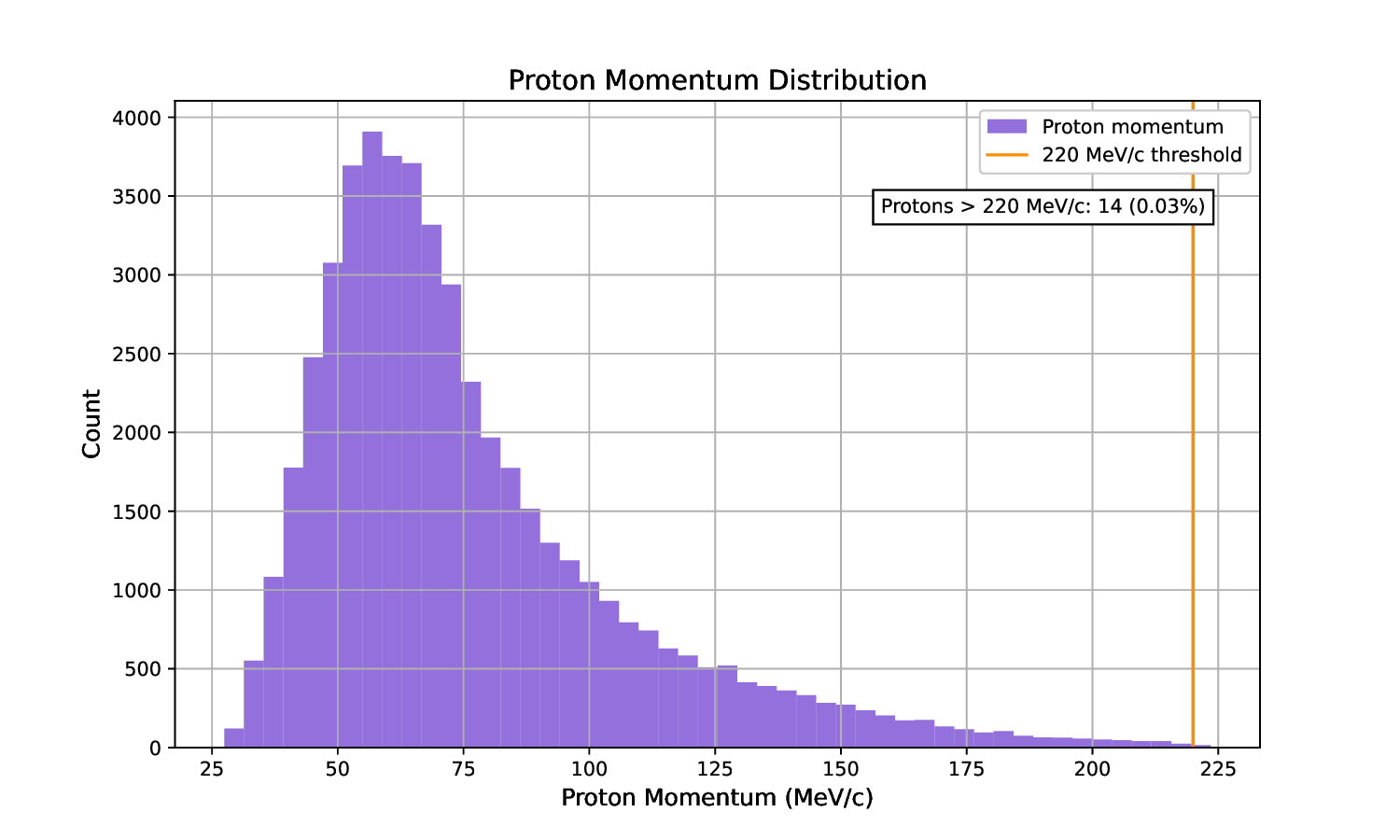} }}%
    \caption{Kinematically allowed momenta for negative pion, proton and helium-4 daughter nucleus in a three-body mesonic decays of $^5_{\Lambda} He$}%
    \label{fig:negative_plots}%
\end{figure}

The density and count plots are shown in Figure \ref{fig:negative_plots}. The results on pion momentum spectral peaks are summarised in Table \ref{tab:summary}. Table \ref{tab:compare} reveals a comparison between our results with those from other studies and experiment. A striking feature of our work is that the pion momentum peaks determined based solely on 4-momentum conservation  are very close to those observed experimentally or in other theoretical studies. In Ref. \cite{stra1993}, the values $p_{\pi^-}=93.6$ MeV/c and $p_{\pi^-}=100$ Mev/c were reported. The first pion momentum corresponds to a situation when $^4\text{He}$ remains at rest and the nucleon and pion go back to back while the second corresponds to the case where the nucleon and $^4\text{He}$ recoil together. In that same study, values of $p_{\pi^0}=99.6$ Mev/c and $p_{\pi^0}=106$ Mev/c were reported. In Refs. \cite{mot1991, mot1992} sharp peaks are reported at $p_{\pi^-}=99.9$ MeV/c and $p_{\pi^0}=105.2$ Mev/c. In Monte Carlo simulations to test the efficiency of detector setup, a value of $p_{\pi^0}=104.9$ Mev/c was assumed in Ref. \cite{oka2005} for the neutral pion decay in $^5 _\Lambda$He. A value of $p_{\pi^-}=99.14$ MeV/c is reported from calculations using $B_\Lambda$ measured through emulsion experiments \cite{chr2013}. These results clearly show that 4-momentum conservation is the principal framework for mesonic decay.

\begin{table}[h]
    \centering
    \caption{Peaks of pion momenta in mesonic decay of $^5 _\Lambda$He from relativistic kinematics.}
    \begin{tabular}{c|c|c|c}\hline
                           & Q-value                            & Two-body decay           & Three-body decay \\ \hline
$\pi^-$ decay & $Q_{\pi^-}^{\text{eff}}=32.76$ MeV & $p_{\pi^-}=99.27$ MeV/c  & $p_{\pi^-}=97.3$ MeV/c   \\ 
$\pi^0$ decay & $Q_{\pi^0}^{\text{eff}}=37.29$ MeV & $p_{\pi^0}=105.12$ MeV/c  & $p_{\pi^0}=103.0$ MeV/c   \\ \hline
    \end{tabular}
    \label{tab:summary}
\end{table}

\begin{table}[h]
    \centering
    \caption{Comparison between our results for peaks of pion momenta in three-body decay mesonic decay with those from theory and experiment.}
    \begin{tabular}{c|c|c|c|c}\hline
& This paper  & Ref. \cite{stra1993}  & Ref. \cite{mot1991, mot1992} & Ref. \cite{chr2013} (Exp.)\\ \hline
$p_{\pi^-}$/MeV/c & 97.3            & 93.6/100              &  99.9                 & 99.14  \\ 
$p_{\pi^0}$MeV/c  & 103.0          & 99.6/106               &  105.2                & --  \\ \hline
\end{tabular}
\label{tab:compare}
\end{table}

Right after a mesonic decay, the decay products may undergo interactions between themselves while still inside the nucleus or before they reach the detector. These final-state interactions (FSIs) usually distort the momentum distributions presented in this paper \cite{har1968, rao1970}. Mesonic decay itself involves only the weak force. However, some FSIs may involve the nuclear force. For example, in the case of three-body negative pion mesonic decay, three FSIs can modify the momentum distribution through various nuclear reactions, as listed here. 

\begin{enumerate}[label=(\arabic*)]
    \item $\pi^--^4\text{He}$ interactions: Pion absorption ($\pi^- + ^4 \text{He} \to \,\,\, ^3\text{H} + n$), elastic scattering ($\pi^- + ^4\text{He} \to \pi^- + ^4\text{He}$), inelastic scattering  ($\pi^- + ^4\text{He} \to \pi^- + ^4\text{He}^*$) and pion charge exchange ($\pi^- + ^4\text{He} \to ^4\text{H} + \pi^0$ or $\pi^- + ^4\text{He} \to ^3\text{H} + \pi^0$).  
    \item $\pi^--p$ interactions: Elastic scattering ($\pi^-+p \to \pi^-+p$ ), pion charge exchange ($\pi^-+p \to n + \pi^0$), formation of $\Delta^0$ resonances ($\pi^-+p \to \Delta^0$)
    \item $p-^4\text{He}$ interactions: Elastic rescattering ($p+^4\text{He} \to p+^4\text{He}$), proton absorption ($p+^4\text{He} \to ^5\text{Li}$). The p-wave resonances ($p_{1/2}$ and $p_{3/2}$) in the $p-^4\text{He}$ subsystem are well-known features of this mesonic decay \cite{hil1958, tan1958, bye1959, amm1959, kum1996}.   
\end{enumerate}

Some of these nuclear reactions require high threshold momenta before they can occur. From the simulations carried out in this paper through 4-momentum conservation, the range of $\pi^-$, proton and 4He momenta are

\begin{enumerate}[label=(\alph*)]
    \item $\pi^-$: 5.97 to 99.20 MeV/c (Average = 92.84 MeV/c)
    \item p: 27.46 to 223.50 MeV/c (Average = 76.18 MeV/c)
\end{enumerate}

For the case of neutral pion decay, the range of momenta are:

\begin{enumerate}[label=(\alph*)]
    \item $\pi^0$: 6.38 to 105.05  MeV/c (Average= 98.34 MeV/c)
    \item n: 29.08 to 238.75 MeV/c (Average = 80.71 MeV/c)
\end{enumerate}

By observing these momenta ranges, one may immediately conclude that some of these reactions would be impossible for the mesonic decay $^5 _\Lambda \text{He}$. Secondly, in some cases where the threshold momenta is available for the nuclear reaction, the number of events would be very few, making that effect of that reaction negligible. Therefore, only a few of these nuclear reactions actually affect the momentum distribution, and sometimes they do so only negligibly. Nuclear reactions that are very significant even at low momenta include (ii) $\pi^--^4\text{He}$ interaction: elastic scattering and pion absorption (facilitated by Coulomb attraction) (ii) $\pi^--p$ interactions: elastic scattering (iii) $p-4He$ interaction: elastic scattering. In the case of the $p-^4\text{He}$ interaction, Coulomb repulsion prevents absorption at low momenta. Therefore, 4-momentum conservation sets the principal framework as it sets the final-state momenta that determine which nuclear reactions will take place. 

An analysis based on 4-momentum conservation is also useful for planned experiments to guide the setup of equipment. For example, it may provide information to optimise detector acceptance or spectrometer settings. The computation of momentum distributions is important not just for mesonic decays, but for any type of three-body decay. For example, in Ref. \cite{gar2005} the hyperspherical harmonics method is applied for the three-body decay of the $2^+$ state of $^6\text{He}$ and in Refs. \cite{alv2007, alv2009} the R-matrix and Faddeev methods are applied in the study of the $0^+$ and $1^+$ states of $^{12}\text{C}$.

\subsection{Pauli blocking}

The proton and neutron Fermi momenta of 4He are $p^p_F=p^n_F=220$ MeV/c \cite{hyd2014}. All nuclear states, from low-momenta up to $p^p_F$ and $p^n_F$, are already occupied. Our results on the three-body decay shows that almost all the neutrons and protons are Pauli blocked since they have momenta less than the Fermi momentum. Only 0.27\% of neutral-pion decay events and 0.03\% of negative-pion decay events are Pauli allowed.  Therefore, roughly 9 times more neutral-pion decay events are Pauli allowed than negative-pion decay events. Within our study, the fact that neutral pion decay suffers less Pauli blocking can be attributed to the higher Q-value.

\section{Conclusion}

In this paper, we have demonstrated that four-momentum conservation serves as the principal framework for describing the mesonic decay of the \(^{5}_{\Lambda}\text{He}\) hypernucleus. By applying relativistic kinematics within both two-body and three-body decay channels, we have systematically derived the momentum distributions of the final-state particles without invoking dynamical assumptions beyond conservation laws.

For the fully constrained two-body decays, the \(\pi^{-}\) and \(\pi^{0}\) momenta are monochromatic, with values of 99.27\,MeV/\(c\) and 105.12\,MeV/\(c\), respectively, each collinear with the recoiling daughter nucleus. In the more realistic three-body treatment, where the unstable \(^{5}\text{Li}\) and \(^{5}\text{He}\) fragments break up into \(^{4}\text{He} + p\) or \(^{4}\text{He} + n\), the conservation of four-momentum alone defines a kinematical phase space that we sampled via a Monte Carlo approach. The resulting pion momentum distributions peak at 97.3\,MeV/\(c\) for \(\pi^{-}\) and 103.0\,MeV/\(c\) for \(\pi^{0}\), in excellent agreement with previous experimental and theoretical results. This close correspondence underscores that relativistic kinematics, rather than detailed final-state interaction models, provides the dominant constraint on the decay momenta.

Furthermore, our analysis of Pauli blocking, based on the Fermi momentum of \(^{4}\text{He}\), reveals that only \(0.27\%\) of neutral-pion decays and \(0.03\%\) of negative-pion decays are Pauli-allowed. Finally, by computing the full range of kinematically allowed momenta for each decay product, we establish a quantitative basis for identifying which final-state nuclear reactions are energetically possible. Thus, four-momentum conservation not only reproduces key observables but also provides a predictive tool for experimental design and for assessing the role of final-state interactions in light hypernuclei. 

\section*{Data Availability Statement}
Data sharing is not applicable to this article as no new data were created or analyzed in this study.

\nocite{*}
\bibliography{aipsamp}

@article{dal1959,
  title = {Pionic Decay Modes of Light $\ensuremath{\Lambda}$ Hypernuclei},
  author = {Dalitz, R. H. and Liu, L.},
  journal = {Phys. Rev.},
  volume = {116},
  issue = {5},
  pages = {1312--1321},
  numpages = {0},
  year = {1959},
  month = {Dec},
  publisher = {American Physical Society},
  doi = {10.1103/PhysRev.116.1312},
  url = {https://link.aps.org/doi/10.1103/PhysRev.116.1312}
}

@article{szy1958,
  author = {J. Szymański},
  title = {Hyperfragment decay energy and angular distribution -- {I}},
  journal = {Il Nuovo Cimento},
  volume = {10},
  number = {5},
  pages = {834--843},
  year = {1958},
  doi = {10.1007/BF02859539}
}

@article{tan1958,
  author = {Y. C. Tang},
  title = {The effect of the final state interaction on the decay of hyper-{${}^5$He}},
  journal = {Il Nuovo Cimento},
  volume = {10},
  number = {5},
  pages = {780--788},
  year = {1958},
  doi = {10.1007/BF02859534}
}

@article{hil1958,
  author = {R. D. Hill},
  title = {Decay of hyper-{${}^5$He}},
  journal = {Il Nuovo Cimento},
  volume = {8},
  number = {3},
  pages = {459--462},
  year = {1958},
  doi = {10.1007/BF02828753}
}

@article{gaj1969,
title = {Study of the $\pi^-$ mesonic decay of the $\Lambda$5He hypernucleus},
journal = {Nuclear Physics B},
volume = {14},
number = {1},
pages = {11-27},
year = {1969},
issn = {0550-3213},
doi = {https://doi.org/10.1016/0550-3213(69)90340-X},
url = {https://www.sciencedirect.com/science/article/pii/055032136990340X},
author = {W. Gajewski and J. Naisse and J. Sacton and P. Vilain and G. Wilquet}
}

@article{kum1996,
  author = {Izumi Kumagai-Fuse and Shigeto Okabe and Yoshinori Akaishi},
  title = {Pionic-decay spectra of few-body $\Lambda$ hypernuclei},
  journal = {Physical Review C},
  volume = {54},
  number = {6},
  pages = {2843--2850},
  year = {1996},
  doi = {10.1103/PhysRevC.54.2843}
}

@article{amm1959,
  author = {R. Ammar and R. Levi Setti and W. E. Slater and S. Limentani and P. E. Schlein and P. H. Steinberg},
  title = {Final state interaction in {${}^5\text{He}_\Lambda$} decay},
  journal = {Il Nuovo Cimento},
  volume = {13},
  number = {6},
  pages = {1156--1164},
  year = {1959},
  doi = {10.1007/BF02725126}
}

@article{rao1970,
  author = {N. K. Rao and M. S. Swami and A. Gurtu and M. B. Singh},
  title = {Final state interactions in the decay of light hypernuclei},
  journal = {Il Nuovo Cimento A},
  volume = {71},
  number = {6},
  pages = {257--265},
  year = {1970},
  doi = {10.1007/bf03049572}
}

@article{bye1959,
title = {Energy and angular distributions of mesonic hyperfragment decays},
journal = {Nuclear Physics},
volume = {11},
pages = {554-568},
year = {1959},
issn = {0029-5582},
doi = {https://doi.org/10.1016/0029-5582(59)90297-4},
url = {https://www.sciencedirect.com/science/article/pii/0029558259902974},
author = {N. Byers and W.N. Cottingham}
}

@article{har1968,
title = {Final state interaction effects in $\pi^-$ mesonic decays of light hypernuclei},
journal = {Nuclear Physics B},
volume = {4},
number = {3},
pages = {277-294},
year = {1968},
issn = {0550-3213},
doi = {https://doi.org/10.1016/0550-3213(68)90312-X},
url = {https://www.sciencedirect.com/science/article/pii/055032136890312X},
author = {D.M. Harmsen and R. {Levi Setti} and J. Zakrzewski and W. Gajewski and J. Sacton and P. Vilain and G. Wilquet and D.H. Davis and E.R. Fletcher and J.E. Allen and A.P. Conway}
}

@article{mot1992,
title = {Mesonic weak decays of $\Lambda$ - and $\Lambda \Lambda$-hypernuclei},
journal = {Nuclear Physics A},
volume = {547},
number = {1},
pages = {115-126},
year = {1992},
issn = {0375-9474},
doi = {https://doi.org/10.1016/0375-9474(92)90717-X},
url = {https://www.sciencedirect.com/science/article/pii/037594749290717X},
author = {T. Motoba}
}

@article{alb2002,
title = {Weak decay of $\Lambda$-hypernuclei},
journal = {Physics Reports},
volume = {369},
number = {1},
pages = {1-109},
year = {2002},
issn = {0370-1573},
doi = {https://doi.org/10.1016/S0370-1573(02)00199-0},
url = {https://www.sciencedirect.com/science/article/pii/S0370157302001990},
author = {W.M. Alberico and G. Garbarino}
}

@inproceedings{bur1965,
  author    = {E. H. S. Burhop},
  title     = {Hypernuclei},
  booktitle = {Proceedings of the 1964 Easter School for Physicists},
  year      = {1964},
  editor    = {E.H.S. Burhop and J.C. Combe and C. Franzinetti and L. Van Hove},
  publisher = {CERN},
  address   = {Geneva},
  pages     = {137}
}

@article{wan2021,
doi = "10.1088/1674-1137/abddaf",
url = "https://dx.doi.org/10.1088/1674-1137/abddaf",
year = "2021",
month = "mar",
publisher = "Chinese Physical Society and the Institute of High Energy Physics of the Chinese Academy of Sciences and the Institute of Modern Physics of the Chinese Academy of Sciences and IOP Publishing Ltd",
volume = "45",
number = "3",
pages = "030003",
author = "Wang, Meng and Huang, W.J. and Kondev, F.G. and Audi, G. and Naimi, S.",
title = "The AME 2020 atomic mass evaluation (II). Tables, graphs and references*",
journal = "Chinese Physics C"
}

@article{stra1993,
title = "Mesonic decay of 5$\Lambda$He with quark-model-based hypernuclear wave function",
journal = "Nuclear Physics A",
volume = "556",
number = "4",
pages = "531-551",
year = "1993",
issn = "0375-9474",
doi = "https://doi.org/10.1016/0375-9474(93)90469-E",
url = "https://www.sciencedirect.com/science/article/pii/037594749390469E",
author = "U. Straub and J. Nieves and A. Faessler and E. Oset"
}

@article{maj2006,
    author = "Majlingova, Olga and Sopko, Vit",
    title = "Kinematic Analysis of Relativistic Hypernuclei Decays",
    journal = "AIP Conference Proceedings",
    volume = "831",
    number = "1",
    pages = "496-498",
    year = "2006",
    month = "04",
    issn = "0094-243X",
    doi = "10.1063/1.2200994",
    url = "https://doi.org/10.1063/1.2200994",
    eprint = "https://pubs.aip.org/aip/acp/article-pdf/831/1/496/11689668/496\_1\_online.pdf",
}

@article{mot1991,
title = "Continuum pion spectra in the weak decays of $\Lambda$4H, $\Lambda$5He and $\Lambda \Lambda$6He",
journal = "Nuclear Physics A",
volume = "534",
number = "3",
pages = "597-619",
year = "1991",
issn = "0375-9474",
doi = "https://doi.org/10.1016/0375-9474(91)90463-G",
url = "https://www.sciencedirect.com/science/article/pii/037594749190463G",
author = "T. Motoba and H. Bandō and T. Fukuda and J. Žofka"
}

@article{oka2005,
title = "$\pi^0$ decay branching ratios of 5$\Lambda$He and 12$\Lambda$C hypernuclei",
journal = "Nuclear Physics A",
volume = "754",
pages = "178-183",
year = "2005",
note = "Proceedings of the Eighth International Conference on Hypernuclear and Strange Particle Physics",
issn = "0375-9474",
doi = "https://doi.org/10.1016/j.nuclphysa.2005.02.128",
url = "https://www.sciencedirect.com/science/article/pii/S0375947405002757",
author = "S. Okada and S. Ajimura and K. Aoki and A. Banu and H.C. Bhang and T. Fukuda and O. Hashimoto and J.I. Hwang and S. Kameoka and B.H. Kang and E.H. Kim and J.H. Kim and M.J. Kim and T. Maruta and Y. Miura and Y. Miyake and T. Nagae and M. Nakamura and S.N. Nakamura and H. Noumi and Y. Okayasu and H. Outa and H. Park and P.K. Saha and Y. Sato and M. Sekimoto and T. Takahashi and H. Tamura and K. Tanida and A. Toyoda and K. Tsukada and T. Watanabe and H.J. Yim"
}

@article{chr2013,
title = "Study of Light Hypernuclei by Pionic Decay at JLab",
journal = "PR12-10-001",
year = "2013",
url = "https://www.jlab.org/exp_prog/proposals/10/PR12-10-001.pdf",
author = "M. Christy and others"
}

@article{nagao2023,
title = "High-resolution spectroscopy of light hypernuclei with the decay-pion spectroscopy",
journal = "LOI12-23-011",
year = "2023",
url = "https://www.jlab.org/exp_prog/proposals/23/LOI12-23-011.pdf",
author = "Nagao, S. and others"
}

@article{hyd2014,
title = "Proton and Neutron Momentum Distributions in A = 3 Asymmetric Nuclei",
journal = "PR12-13-012",
year = "2014",
url = "https://www.jlab.org/exp_prog/proposals/14/PR12-14-011.pdf",
author = "C. Hyde and others"
}

@article{ess2015,
  title = {Observation of $_{\mathrm{\Lambda}}^{4}\mathrm{H}$ Hyperhydrogen by Decay-Pion Spectroscopy in Electron Scattering},
  author = {Esser, A. and Nagao, S. and Schulz, F. and Achenbach, P. and Ayerbe Gayoso, C. and B{\"o}hm, R. and Borodina, O. and Bosnar, D. and Bozkurt, V. and Debenjak, L. and Distler, M. O. and Fri{\v{s}}{\v{c}}i{\'c}, I. and Fujii, Y. and Gogami, T. and Hashimoto, O. and Hirose, S. and Kanda, H. and Kaneta, M. and Kim, E. and Kohl, Y. and Kusaka, J. and Margaryan, A. and Merkel, H. and Mihovilovi{\v{c}}, M. and M{\"u}ller, U. and Nakamura, S. N. and Pochodzalla, J. and Rappold, C. and Reinhold, J. and Saito, T. R. and Sanchez Lorente, A. and S{\'a}nchez Majos, S. and Schlimme, B. S. and Schoth, M. and Sfienti, C. and {\v{S}}irca, S. and Tang, L. and Thiel, M. and Tsukada, K. and Weber, A. and Yoshida, K.},
  collaboration = {A1 Collaboration},
  journal = {Phys. Rev. Lett.},
  volume = {114},
  number = {23},
  pages = {232501},
  numpages = {5},
  year = {2015},
  month = jun,
  publisher = {American Physical Society},
  doi = {10.1103/PhysRevLett.114.232501},
  url = {https://link.aps.org/doi/10.1103/PhysRevLett.114.232501}
}

@article{ach2022,
title = {Hypertriton Production in $p$-Pb Collisions at $\sqrt{{s}_{NN}}=5.02\text{ }\text{ }\mathrm{TeV}$},
  author = "Acharya, S. and others",
  collaboration = "A Large Ion Collider Experiment Collaboration",
  journal = "Phys. Rev. Lett.",
  volume = "128",
  issue = "25",
  pages = "252003",
  numpages = "13",
  year = "2022",
  month = "Jun",
  publisher = "American Physical Society",
  doi = "10.1103/PhysRevLett.128.252003",
  url = "https://link.aps.org/doi/10.1103/PhysRevLett.128.252003"
}

@article{gar2005,
title = "Anatomy of three-body decay III: Energy distributions",
journal = "Nuclear Physics A",
volume = "766",
pages = "74-96",
year = "2006",
issn = "0375-9474",
doi = "https://doi.org/10.1016/j.nuclphysa.2005.12.001",
url = "https://www.sciencedirect.com/science/article/pii/S0375947405012406",
author = "E. Garrido and D.V. Fedorov and A.S. Jensen and H.O.U. Fynbo"
}

@article{alv2009,
  title = "Three-Body Decays: Structure, Decay Mechanism and Fragment Properties",
  author = {\'Alvarez-Rodr\'{\i}guez, R. and Garrido, E. and Fedorov, D. V. and Fynbo, H. O. U. and Kirsebom, O. S.},
  journal = "Few-Body Systems",
  volume = "45",
  issue = "2",
  pages = "149-152",
  year = "2009",
  doi = "10.1007/s00601-009-0010-2"
}

@article{alv2007,
  title = "Energy Distributions from Three-Body Decaying Many-Body Resonances",
  author = {\'Alvarez-Rodr\'{\i}guez, R. and Jensen, A. S. and Fedorov, D. V. and Fynbo, H. O. U. and Garrido, E.},
  journal = "Phys. Rev. Lett.",
  volume = "99",
  issue = "7",
  pages = "072503",
  numpages = "4",
  year = "2007",
  month = "Aug",
  publisher = "American Physical Society",
  doi = "10.1103/PhysRevLett.99.072503",
  url = "https://link.aps.org/doi/10.1103/PhysRevLett.99.072503"
}

@CONTROL{REVTEX41Control}

@CONTROL{aip41Control,pages="1",title="0"}

\end{document}